\documentclass{elsart}
\usepackage{amsmath}
\usepackage{amssymb}
\usepackage[dvips]{graphicx}
\usepackage[sl,footnotesize]{subfigure}

\newcommand{\ip}{\ensuremath{\text{.}}}
\newcommand{\ic}{\ensuremath{\text{,}}}
\newcommand{\figref}[1]{\mbox{{\slshape\figurename\ \ref{#1}}}{\kern0pt}}
\newcommand{\tabref}[1]{\mbox{{\slshape\tablename\ \ref{#1}}}{\kern0pt}}

\newcommand{\stt}[1]{\ensuremath{\mathrm{#1}}}
\newcommand{\unit}[1]{\ensuremath{\,\mathrm{#1}}}
\newcommand{\degree}{\ensuremath{^\circ}}

\newcommand{\phix}{\ensuremath \phi_\stt{x}}
\newcommand{\phiy}{\ensuremath \phi_\stt{y}}
\newcommand{\xirf}{\ensuremath \Delta\xi_\stt{fps}}
\newcommand{\etarf}{\ensuremath \Delta\eta_\stt{fps}}
\newcommand{\DC}{Davies--Cotton}

\begin{document}

\begin{frontmatter}

\title{%
Wide-field prime-focus Imaging Atmospheric Cherenkov Telescopes: A systematic study
}

\author{Albert Schliesser\corauthref{cor}},
\corauth[cor]{Corresponding author. Tel./Fax. +49-89-32354-328}
\ead{aschlies@mppmu.mpg.de}
\author{Razmick Mirzoyan}
\address{Max-Planck-Institut f\"ur Physik, F\"ohringer Ring 6, D-80805 M\"unchen, Germany}

\begin{abstract}
By means of third-order optical theory as well as ray-tracing
simulations we have investigated the feasibility of wide-field
imaging atmospheric {Cherenkov} telescopes
with a reflective prime-focus design.
For a range of desired optical resolutions, we have determined
the largest available field-of-view of single-piece spherical,
single-piece parabolic, tessellated spherical, tessellated
para\-bolic and \DC{} designs, always considering a wide range of
design parameters.
The \DC{} design exhibits a surprising similarity to the
tessellated parabolic design in its qualitative
behaviour.
Also, elliptic telescope designs with better off-axis imaging
properties than \DC{} are presented.
We show that by using $f/2$ 
optics it is possible to build prime-focus telescopes
with a full field-of-view of $10\degree$ at $0{.}1\degree$
resolution.

\end{abstract}

\end{frontmatter}

\bibliography{magic}
\bibliographystyle{elsart-num}

\section{Introduction}\label{s:int}
The first strong $\gamma$-ray signal in the TeV energy range
has been measured
by the Whipple collaboration from the Crab Nebula in 1989 \cite{Weekes1989}.
Since then the technique of ground-based
imaging atmospheric Cherenkov telescopes (IACTs) has
substantially improved.
Nowadays the ground-based IACT
technique is the most sensitive method for measuring
very high energy (VHE) $\gamma$-rays from celestial objects.
%
%
The CANGAROO \cite{Kawachi2001}, HESS \cite{Bernlohr2003},
MAGIC \cite{Fernandez1998} and VERITAS \cite{Weekes2002}
telescopes are currently the largest and the most
sensitive instruments measuring the sky in $\gamma$-rays from
four continents.
Although differing in detail, their optics are all prime-focus systems
consisting of a single, large-aperture, segmented reflector and
a 2-dimensional detector array in the focal plane.
These systems are delivering highly interesting scientific data,
although some of them are not yet in their final configuration.
In the next 2-3 years they will be completed and put into
operation at their full power.
The community hopes that these improved instruments will increase
the number of established VHE $\gamma$-ray sources by one order of
magnitude.

The evaluation of the scientific outcome might eventually also reveal
which demands the next-generation instrumentation should meet.
In recent years one of the frequently discussed designs are wide-angle
IACTs \cite{Kifune1997}.
By using wide-angle, highly sensitive, large telescopes
of very low threshold energy setting one can perform
all-sky surveys in a short time.
In order to discriminate images induced by $\gamma$-ray showers
from those of much more abundant hadrons an optical resolution of
$\sim 0{.}1\degree$ is required over the entire field-of-view (FOV).
Below 100 GeV, even $\geq 2$ times better
resolution is needed, since the differences in the images of $\gamma$-ray-
and hadron-induced
showers become smaller at lower energies.
Thus, depending on the desired resolution, the FOV of present-day IACTs is
limited to below $5\degree$.

In the following, we present an exhaustive analysis of the common IACT
designs with respect to their wide-field performance.
For that purpose, we have analysed optical spot sizes as a function
of focal ratio, incidence angle and mirror segment size by means of
both analytical third-order geometrical optics and ray-tracing
simulations.

\section{Methods}\label{s:single}

Both approaches used in this paper are capable of determining
the optical spot size: 
By using third-order optical theory, on the one hand, one
can readily assess a given system by evaluating analytically
derived formulae.
Yet, its results are only approximative and the theory can hardly
be applied to tessellated geometries.
Ray-tracing simulations, on the other hand, can predict optical
performance very accurately, but require proper simulation of
every given system and parameter set.


\subsection{Parameters and properties of prime focus systems}

\begin{figure}[p]
  \includegraphics[width=\linewidth]{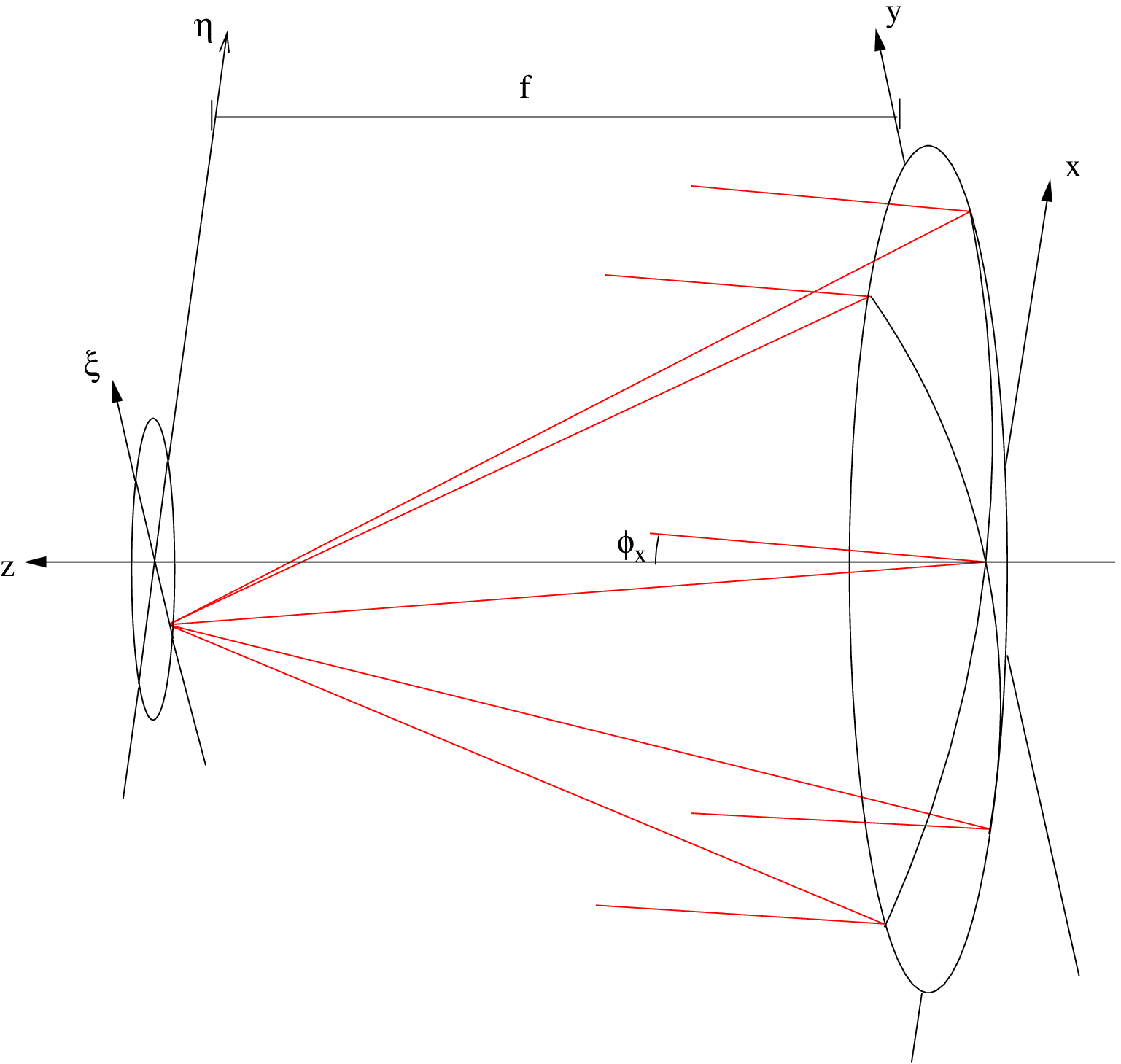}
  \caption{Optical layout of a prime focus system.
    Coming from the left, the rays hit the mirror in coordinates
    $(x,y,z(x,y))$, are reflected and intercept the
    focal plane in $(\xi,\eta,f)$.
    Some examples of off-axis rays (red) with an incidence
    angle $\phix$ are indicated.
  }
  \label{f:pfs}
\end{figure}

The basic geometry of a simple prime focus system is depicted in
\figref{f:pfs}.
Every ray is incident on the reflector in a point $(x,y,z(x,y))$,
is reflected and hits the image plane in $(\xi,\eta, f)$,
where $f$ is the focal length of the system.
Here, $z(x,y)$ is the reflector surface function.
The reflector and image coordinate systems are supposed to be aligned
such that $\vec e_{\xi} \parallel \vec e_{x}$ and
$\vec e_{\eta} \parallel \vec e_{y}$,
but are offset along the global $z$-axis by the
focal distance $f$.
Although in general, the image surface could be curved and offset along
the $z$-axis to correct for aberrations, we will only consider the simpler
case of a fixed and flat image surface.
For IACTs, these are realistic assumptions.

The angular object coordinates $(\phix,\phiy)$ determine
the angles the incident rays make with the $z$-axis.
Single-piece reflectors have full rotational symmetry which is only
slightly disturbed by segmentation to the analyzed extent.
For this reason, an analysis with $\phiy=0$ imposes no restriction, and
we may simply distinguish the {\em tangential\/} image coordinate
(parallel to the projection of the incident ray into the image plane)
and the {\em sagittal\/} image coordinate
(perpendicular to the projection of the incident ray into the image plane):
If $\phix\neq 0= \phiy$, then $\xi$ is the tangential, and $\eta$ the sagittal
image coordinate.

For a given incidence angle $\phix$, rays incident on different reflector loci hit 
the image (camera) plane at slightly different points due to aberrations.
The mean values $(\langle\xi\rangle,\langle\eta\rangle)$
of all incoming rays' image coordinates define the image {\em centroid}.
The {\em rms point spread}, the root-mean-square deviations 
\begin{align}
  \begin{split}
  \Delta\xi &=\sqrt{\left\langle \left(\xi - \langle\xi\rangle\right)^2\right\rangle}
            \qquad\qquad\text{(tangential rms)}\\
  \Delta\eta&=\sqrt{\left\langle \left(\eta - \langle\eta\rangle\right)^2\right\rangle}
            \qquad\qquad\text{(sagittal rms)}
  \end{split}
  \label{e:rmsdef}
\end{align}
of the rays' actual image coordinates from the centroid position,
are a reasonable measure of the optical spot size.

Imaging mirrors commonly are surfaces of revolution
defined by conic sections.
They are characterized by three parameters, namely their
radius of curvature $r$, their diameter $d$ and their conic
\mbox{constant $\delta$}.
Their surface equation reads
\begin{equation}
  \label{e:cs}
  z(x,y)\equiv z(h)=\frac{1}{r}\cdot\frac{h^2}{1+\sqrt{1-(1+\delta)h^2/r^2}}
\end{equation}
with $h:=\sqrt{x^2+y^2}\leq d/2$.
Parabolic and spherical shapes are obtained by setting $\delta=-1$ and
$\delta=0$, respectively.

In the simplest case, the reflector of a telescope consists of one single
large mirror with its surface defined by \eqref{e:cs}.
Though, due to the large apertures and cost reasons, the reflectors of
 existing IACTs are
segmented into considerably smaller mirrors, which are mounted on a
common reflector dish.
The individual mirror segments are conic sections of revolution.
For ease of fabrication and testing, they are mostly chosen to be
spherical.

We assume the individual mirrors of such a tessellated reflector
are square-shaped, and mounted on a
square grid in $(x,y)$ such that the four inmost mirrors touch
$(0,0)$ with one edge.
The $z$-coordinate of the mirror centers is defined
by the {\em gross reflector shape}, which itself is a conic
section of revolution.
The size of the segments is expressed in terms of the
{\em tessellation ratio}
\begin{equation}
  \alpha=\frac{\text{size of individual mirror}}
              {\text{diameter of reflector}}\ip
\end{equation}

Tessellation may introduce further aberrations, since the
actual shape of the reflector deviates from the (putatively) ideal
shape of a single reflector.
Alignment accuracy may also influence the optical quality strongly.
Interestingly, tessellation also introduces new degrees of freedom to the
optical design:
An individual mirror's orientation (i.~e. the normal to its surface)
can be chosen independently of the normal to the gross shape at this
mirror's location.


\subsection{Third-order analysis}\label{sss:toal}

Third-order optical theory has been developed for optical
engineering tasks before the advent of fast computers.
It is capable of readily anticipating optical
performance of a given system and its dependence on
design parameters once an appropriate
formula has been derived from its basic rules.
Although in principle it could also be applied to tessellated
systems, its elegant simplicity would be lost due to the need
of summation over a large number of single mirrors.
For that reason, its application will be limited to single mirror systems
in this study.
Only in the limit of small tessellation ratios, third-order results apply
to tessellated systems as well.

When a ray with angular object coordinates $(\phix,\phiy)$
hits a single-piece reflector
in $(x,y,z(x,y))$, then from third-order aberration theory \cite{Korsch1991},
an approximate expression for the image coordinates
can be derived to be
\begin{equation}
\label{e:cents}
  \begin{split}
  \xi &= -\frac{
	    h^2 ( x (1+ \delta) + 2 f \phix )+
	    4 f( 2 f^2 \phix + ( x + 2 f \phix ) ( x \phix + y \phiy ) )}
	  {8 f^2}\\
  \eta&= -\frac{
	    h^2 ( y (1+ \delta) + 2 f \phiy )+
          4 f( 2 f^2 \phiy + ( y + 2 f \phiy ) ( x \phix + y \phiy ) )}
	   {8 f^2}\ip
  \end{split}
\end{equation}

%

By integrating $x$ and $y$ over a circular aperture with $h\leq d/2$,
analytic expressions for the rms point spread 
\eqref{e:rmsdef} can immediately be given as
\begin{align}
  \begin{split}
     \xirf&=\frac{1}{4} \sqrt{
              \frac{(1+\delta)^2}{2048}            \left(\frac{d}{f}\right)^6+
              \frac{(6+4\delta)\phix^2+\phi^2}{96} \left(\frac{d}{f}\right)^4+
	      \phix^2\phi^2                        \left(\frac{d}{f}\right)^2
            }\\
     \etarf&=\frac{1}{4}\sqrt{
               \frac{(1+\delta)^2}{2048}           \left(\frac{d}{f}\right)^6+
               \frac{(6+4\delta)\phiy^2+\phi^2}{96}\left(\frac{d}{f}\right)^4+
               \phiy^2\phi^2                       \left(\frac{d}{f}\right)^2
           }\ic
  \end{split}
  \label{e:torms}
\end{align}
using the abbreviation $\phi:=\sqrt{\phix^2+\phiy^2}$.
These expressions still simplify considerably if we set
$\phiy=0$ as discussed before, this substitution was deferred
in order to show the symmetry of the system.
The index ``fps'' indicates that the results are expressed in the
focal plane scale,  i.\ e.\ they have been divided by the focal
length of the system to yield an angular quantity.
It will be omitted for the sake of
shortness in some places;
generally, if the rms point spread is given in angular units, the focal
plane scaling applies.
The expressions \eqref{e:torms} immediately give performance estimates
for all conceivable single-piece reflectors by simply
plugging in the corresponding parameters characterizing the system.

\subsection{Ray tracing}
Complementary to the analytic, yet approximative third-order
analysis, ray tracing simulations yield precise performance data
for imaging systems, including also the more complex tessellated
reflector geometries.

Both, a commercially available optical simulation package
\cite{ZEMAX} and a self-imple\-men\-ted ray-tracer were used.
With its intuitive graphical user interface and its powerful
analysis tools, the commercial package
allows one comfortable editing and detailed
evaluation of the systems under consideration.
Though, simulations of tessellated reflectors with their large number of
single optical elements\footnote{Simulations with up to $>100000$ single
mirrors have been made.}
 are cumbersome to implement and require very long simulation run times.
For that reason, a custom ray-tracing engine was programmed.
The 
parallelized C-code
can be run as a stand-alone simulation or used from within the commercial
program to enable the combination of tessellated reflectors with
other optical elements.
Arbitrary tessellated reflector geometries are supported and 
easy to set up.

\section{Results}

Comprehensive simulations have been made for the cases of
(1.1) single-piece spherical,
(1.2) single-piece parabolic,
(2.1) tessellated spherical,
(2.2) tessellated parabolic design with constant radii of curvature,
(2.3) \DC{} and
(2.4) tessellated parabolic design with adjusted radii of curvature.
For every design, systems with focal lengths $f$ ranging from 1.0 to
2.9 in steps of 0.1 were investigated.
Since the diameter of the reflector was set to 1 in all systems,
{\em focal ratios\/} $f/d$ equally range from 1.0 to 2.9.
For tessellated systems, the tessellation ratio was varied from 0.005 to 0.080
in steps of 0.005.
Image quality was analyzed for incidence angles between $0{.}0\degree$
and $5{.}8 \degree$ in steps of $0{.}2\degree$
by means of tracing $\sim 800000$ rays through the system and
determining the rms point spread \eqref{e:rmsdef} of the resulting images.
The simulation results are presented in the following section
together with some implications from third-order theory.

\subsection{Single-piece reflectors}

\subsubsection{Single-piece sphere}

\begin{figure}[p]
  \centering
  \subfigure[
    tangential rms
  ]{
    \label{f:ss-rms-sag}
    \includegraphics[width=0.47\linewidth]{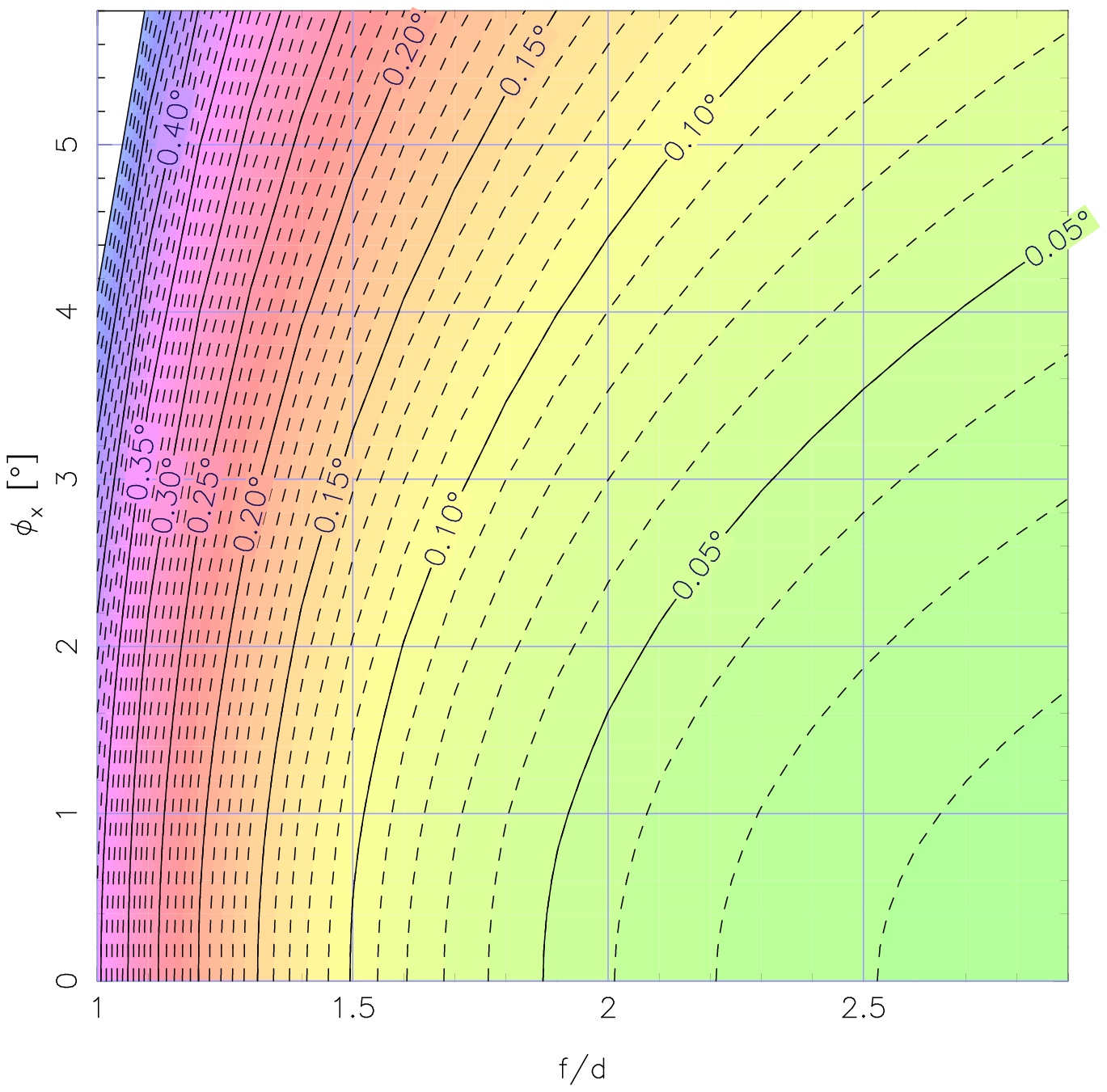}
  }
  \subfigure[
    sagittal rms
  ]{
    \label{f:ss-rms-tan}
    \includegraphics[width=0.47\linewidth]{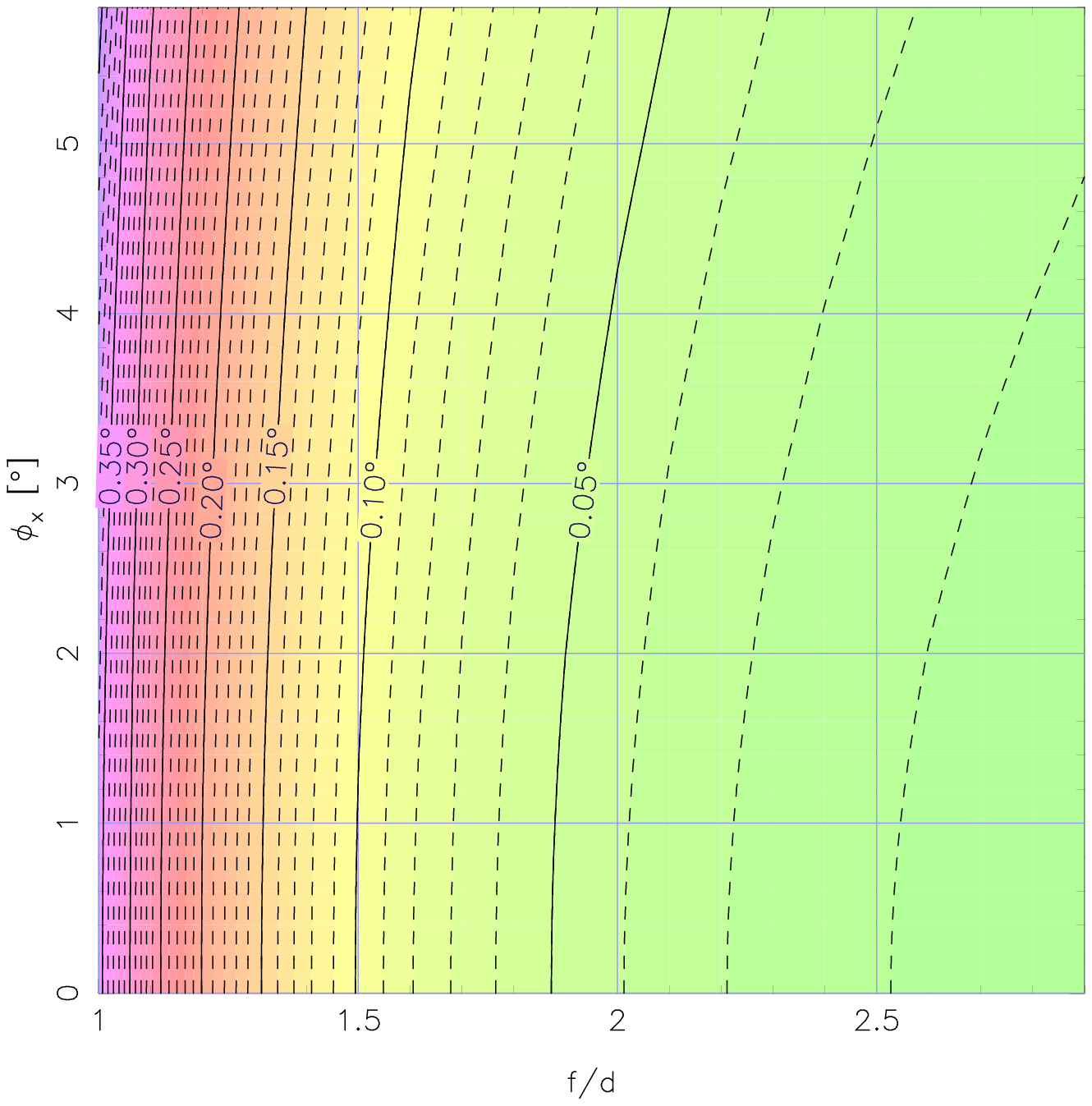}
  }
  \caption{Dependence of rms point spread
           on the incidence angle $\phix$ and the focal ratio $f/d$
	   for a single-piece spherical reflector.
	   The contour lines connect $(f/d,\phix)$-combination which
	   result in an equal rms spread.
	   The line at $0{.}05\degree$ marks the upper limit 
	   for achieving $\gamma$-hadron discrimination.
	   }
  \label{f:ss-rms}
\end{figure}

Inserting $\delta=0$ for a spherical reflector into \eqref{e:torms}
and setting $\phiy=0$ one obtains
\begin{align}
  \begin{split}
     \xirf&=\frac{1}{4} \sqrt{
              \frac{1}{2048}            \left(\frac{d}{f}\right)^6+
              \frac{7\phix^2}{96} \left(\frac{d}{f}\right)^4+
	      \phix^4                        \left(\frac{d}{f}\right)^2
            }\\
     \etarf&=\frac{1}{4}\sqrt{
               \frac{1}{2048}           \left(\frac{d}{f}\right)^6+
               \frac{\phix^2}{96}\left(\frac{d}{f}\right)^4
           }
  \end{split}
  \label{e:tormss}
\end{align}
for the rms point spread in tangential and sagittal direction, respectively.
 {\em Spherical aberration}, corresponding to the first summand in
\eqref{e:tormss}, strongly deteriorates imaging quality
especially for small incidence angles. 
In order to get $\xirf$ and $\etarf$ below $0{.}05 \degree$,
the focal ratio must be larger than $1{.}85$,
as can be derived from \eqref{e:tormss}.
This is confirmed  by the simulation results, which
are shown in \figref{f:ss-rms}.
The sagittal rms hardly changes with the light incidence angle
as it is dominated by spherical aberration in the depicted 
parameter range.
In contrast, the tangential rms displays also a
considerable angular dependence.
The overall behaviour of the system is predicted by third-order
theory with quantitative deviations smaller than $10\%$.

\subsubsection{Single-piece paraboloid}

\begin{figure}[p]
  \centering
  \subfigure[
    tangential rms
  ]{
    \label{f:sp-rms-sag}
    \includegraphics[width=0.47\linewidth]{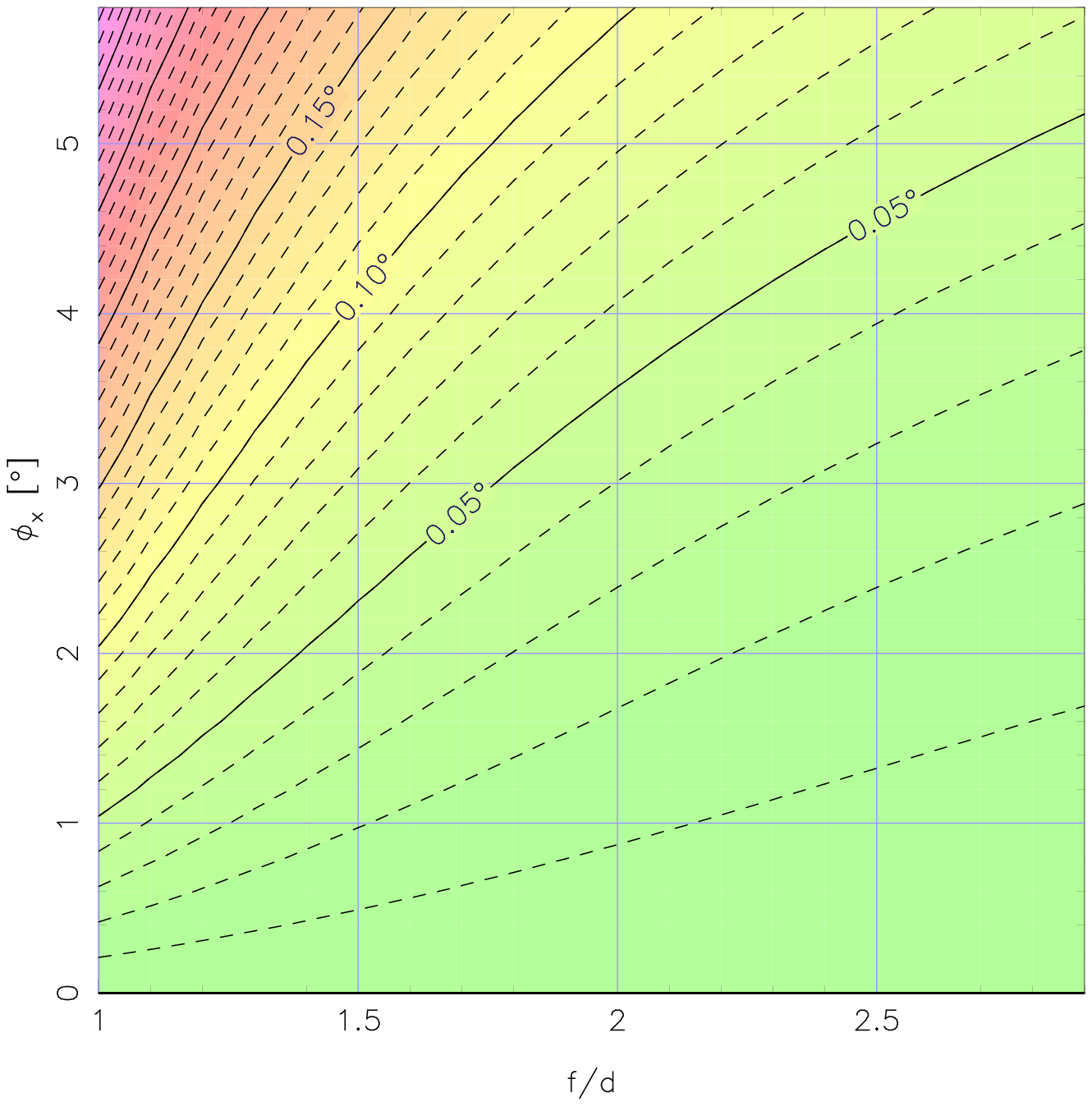}
  }
  \subfigure[
    sagittal rms
  ]{
    \label{f:sp-rms-tan}
    \includegraphics[width=0.47\linewidth]{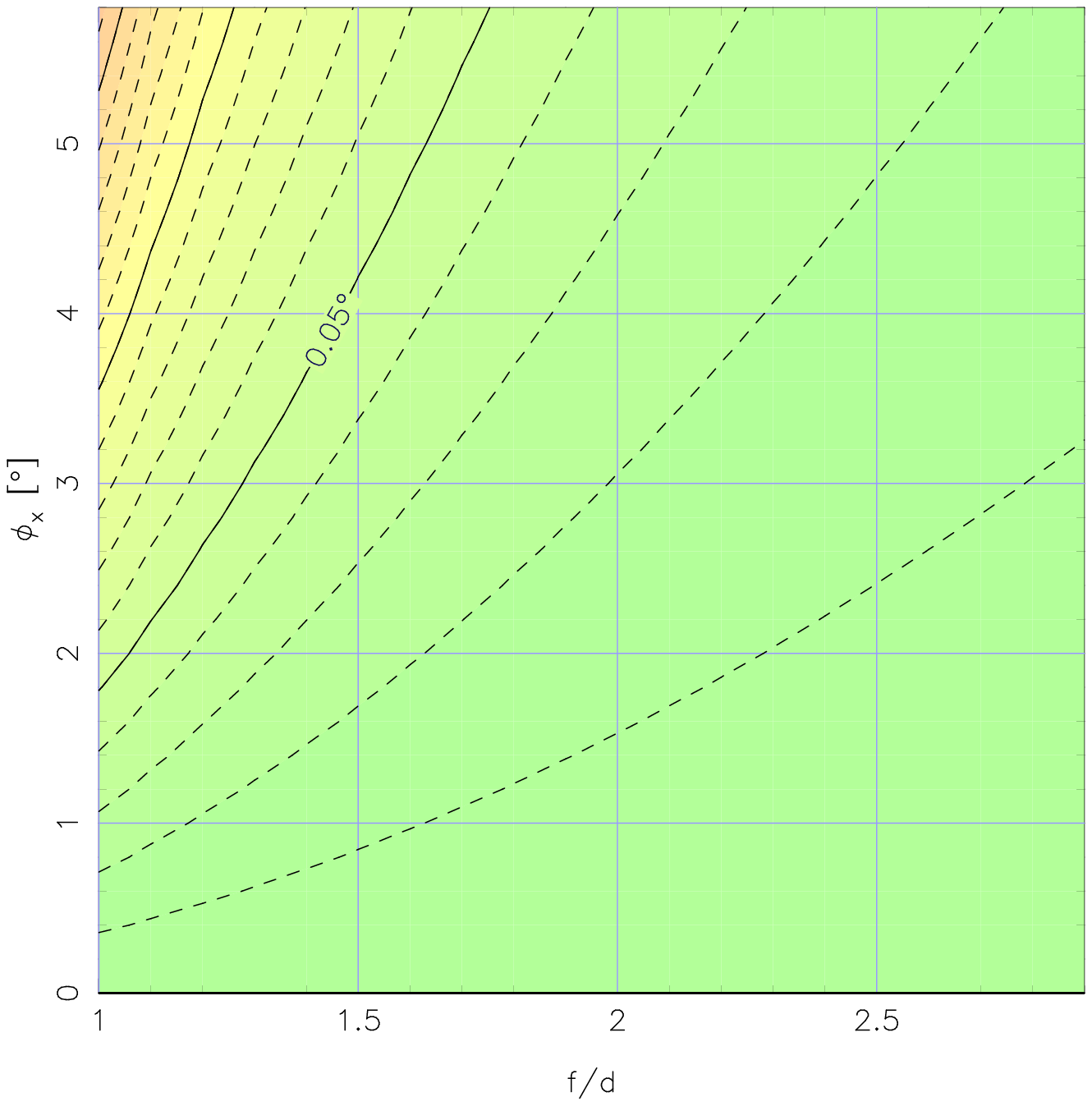}
  }
   
  \caption{Rms point spread for a single-piece parabolic reflector.
	 Illustration analogous to \figref{f:ss-rms}.}
  \label{f:sp-rms}
\end{figure}

Let us now insert the value $\delta=-1$ for a single-piece parabolic
reflector into \eqref{e:torms} and consider again the case $\phiy=0$:
\begin{align}
  \begin{split}
     \xirf&=\frac{1}{4} \sqrt{
              \frac{3\phix^2}{96} \left(\frac{d}{f}\right)^4+
	      \phix^4                        \left(\frac{d}{f}\right)^2
            }\\
     \etarf&=\frac{1}{4}\sqrt{
               \frac{\phix^2}{96}\left(\frac{d}{f}\right)^4
           }
  \end{split}
  \label{e:tormsp}
\end{align}
From the obtained formulae \eqref{e:tormsp} one can notice that
the first summands in the square roots of \eqref{e:torms} have vanished
so that on-axis imaging is supposed to be perfect for arbitrary focal ratios.
Though, at non-zero incidence angles, the aberrations
described by the second ({\em coma}) and third ({\em astigmatism})
summands induce a blurring of the image, which increases progressively
with the incidence angle.
The simulation results, illustrated in \figref{f:sp-rms},
confirm the predictions.

\subsection{Tessellated reflectors}

\subsubsection{Tessellated spherical design}
\begin{figure}[p]
  \centering
  \subfigure[
    tangential rms
  ]{
    \label{f:s-rms-sag}
    \includegraphics[width=0.47\linewidth]{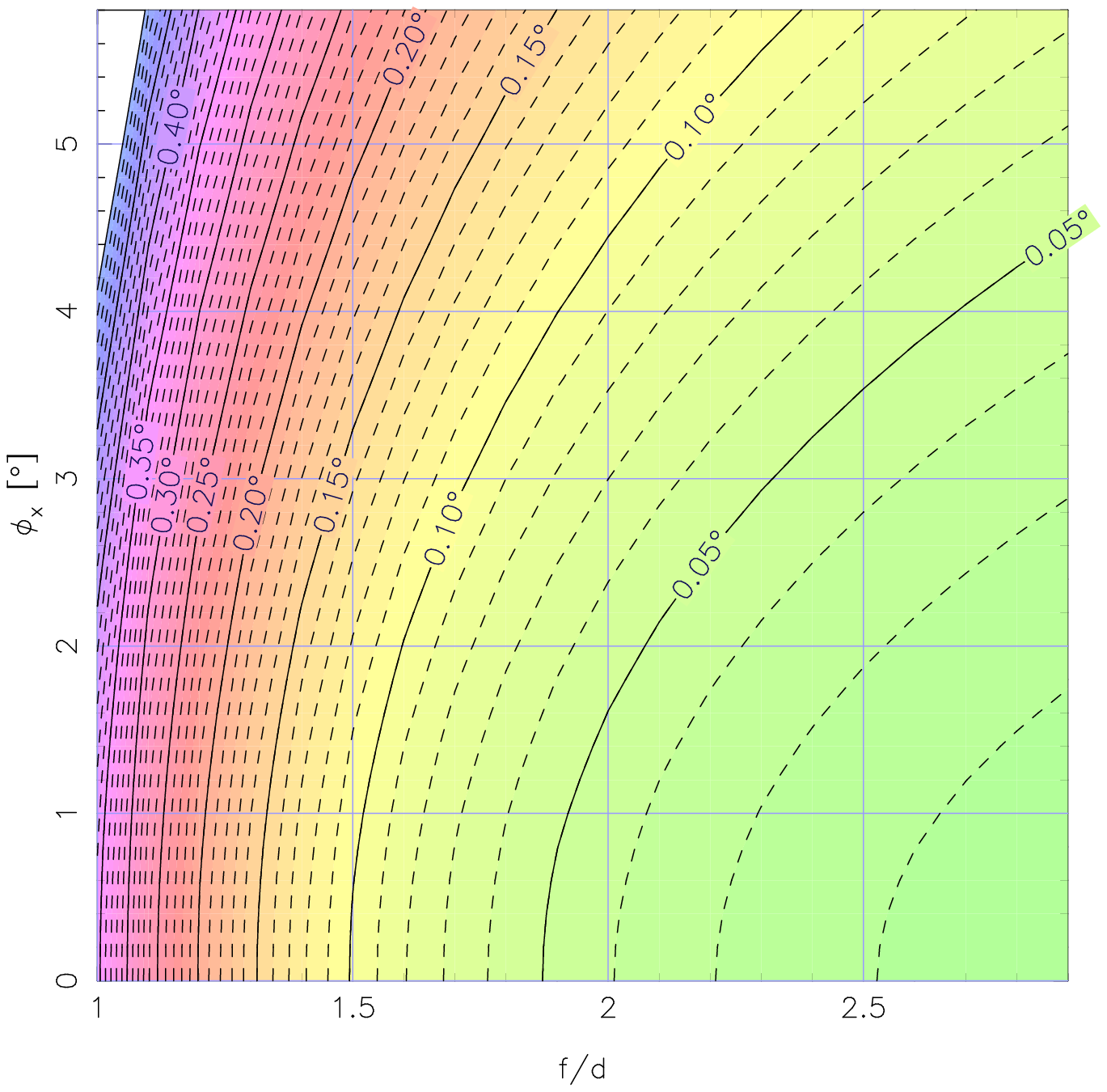}
  }
  \subfigure[
    sagittal rms
  ]{
    \label{f:s-rms-tan}
    \includegraphics[width=0.47\linewidth]{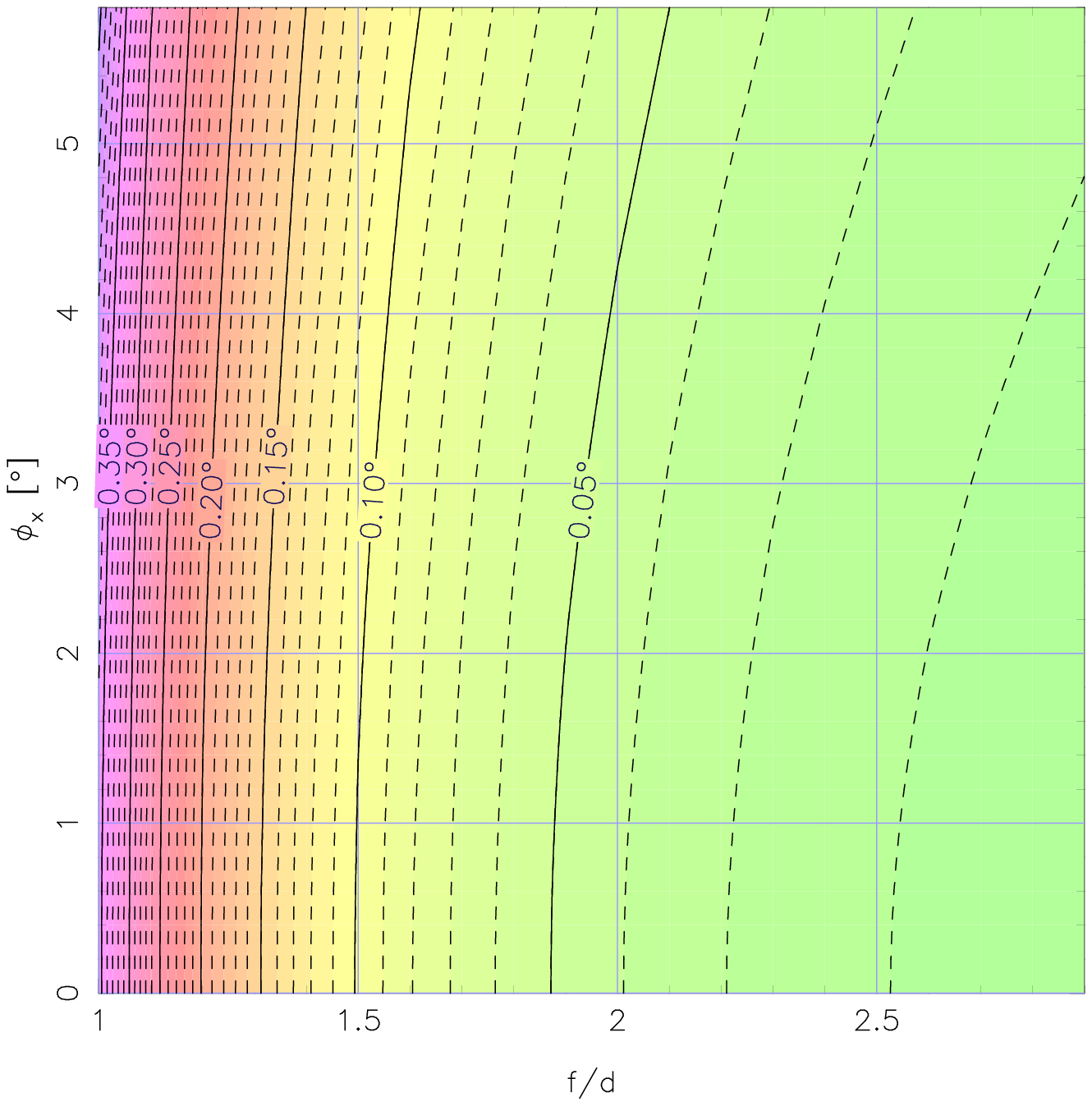}
  }
  \caption{Rms point spread for a tessellated spherical reflector.
  The tessellation ratio $\alpha$ is 0.03.
  Illustration analogous to \figref{f:ss-rms}.}
  \label{f:s-rms}
\end{figure}

A single spherical reflector can be segmented 
into smaller mirrors.
If the small mirrors are spherical themselves and have the same
radius of curvature as the gross sphere
their surfaces coincide with the gross shape.
Apart from possible small gaps between the individual mirrors, 
the tessellated reflector surface is then identical
to the single-piece spherical reflector's.
Correspondingly, the simulation yields also the same performance
data, as illustrated in  \figref{f:s-rms}.

\subsubsection{Tessellated parabolic design with constant radii of
curvature}
\begin{figure}[p]
  \centering
  \subfigure[
    tangential rms
  ]{
    \label{f:p-rms-sag}
    \includegraphics[width=0.47\linewidth]{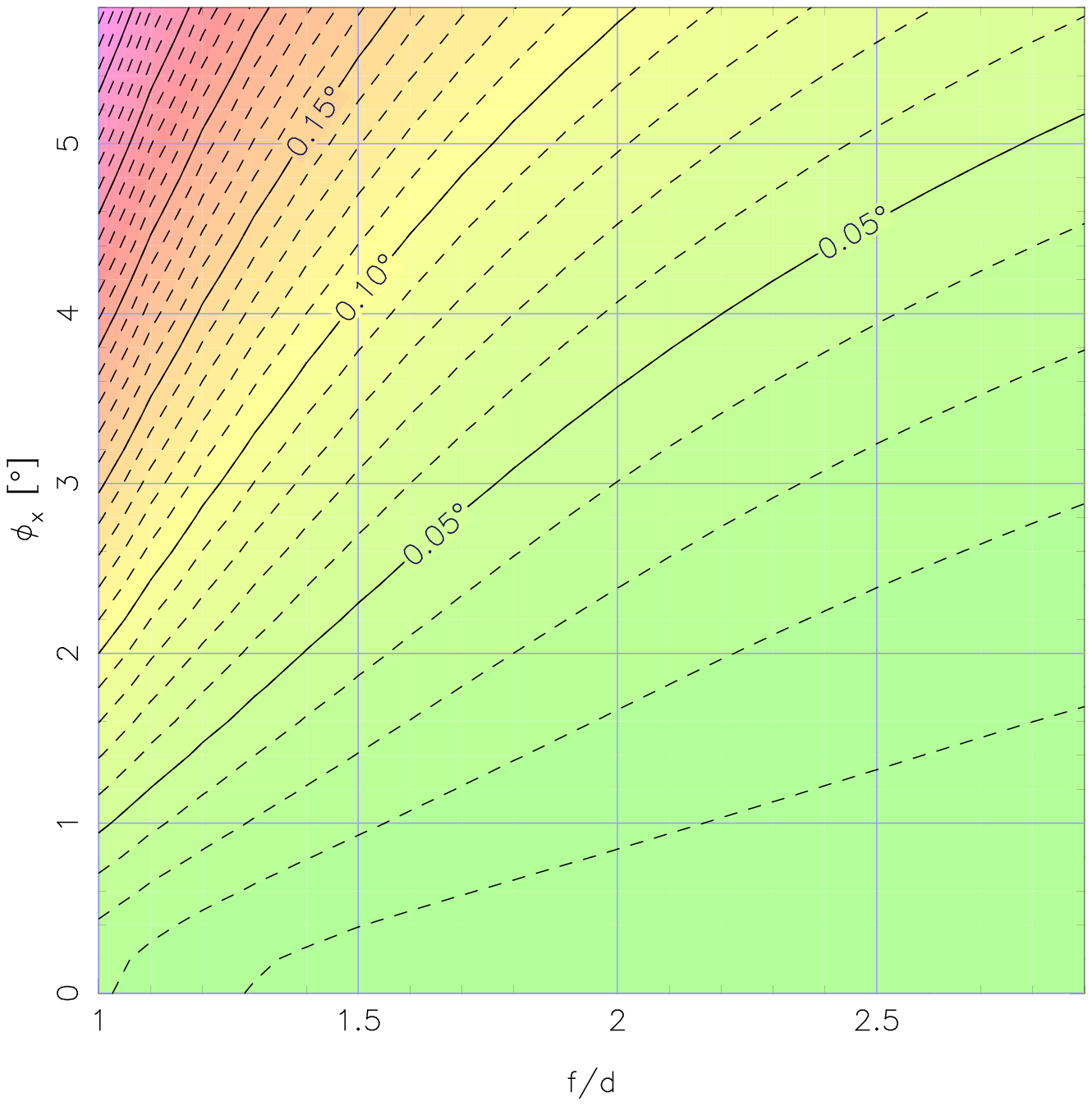}
  }
  \subfigure[
    sagittal rms
  ]{
    \label{f:p-rms-tan}
    \includegraphics[width=0.47\linewidth]{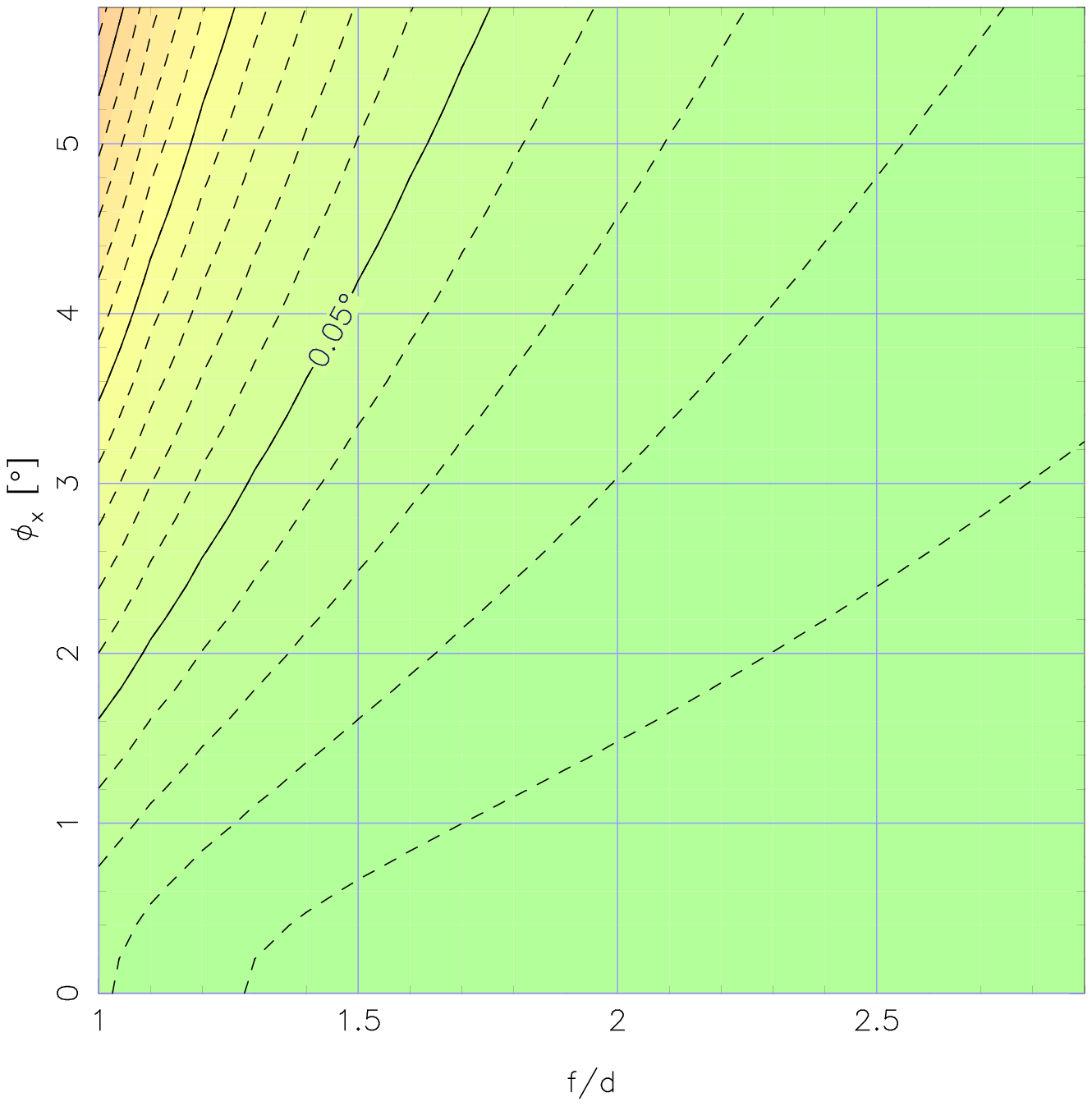}
  }
  \caption{Rms point spread for a tessellated parabolic reflector with constant
           radii of curvature.
	   The tessellation ratio $\alpha$ is 0.03.
            Illustration analogous to \figref{f:ss-rms}.}
  \label{f:p-rms}
\end{figure}

The simplest way to segment a parabolic reflector into smaller mirrors
is the following:
The individual mirrors are spherical, have all the same radius of curvature,
namely twice the focal length $f$ of the telescope, and their 
normals (in the center) coincide with the normal of the gross
reflector shape at their center.
\figref{f:p-rms} shows the simulation results for such a
configuration for a realistic tessellation ratio $\alpha=0{.}03$.
Qualitatively, they are very similar to the results for the
single piece paraboloid. 
Only when approaching $\phix=0$,  where imaging of a single-piece
parabola of revolution becomes perfect, the influence
of tessellation reduces image quality. 

\subsubsection{\DC{} design}
\begin{figure}[p]
  \centering
  \subfigure[
    tangential rms
  ]{
    \label{f:d-rms-sag}
    \includegraphics[width=0.47\linewidth]{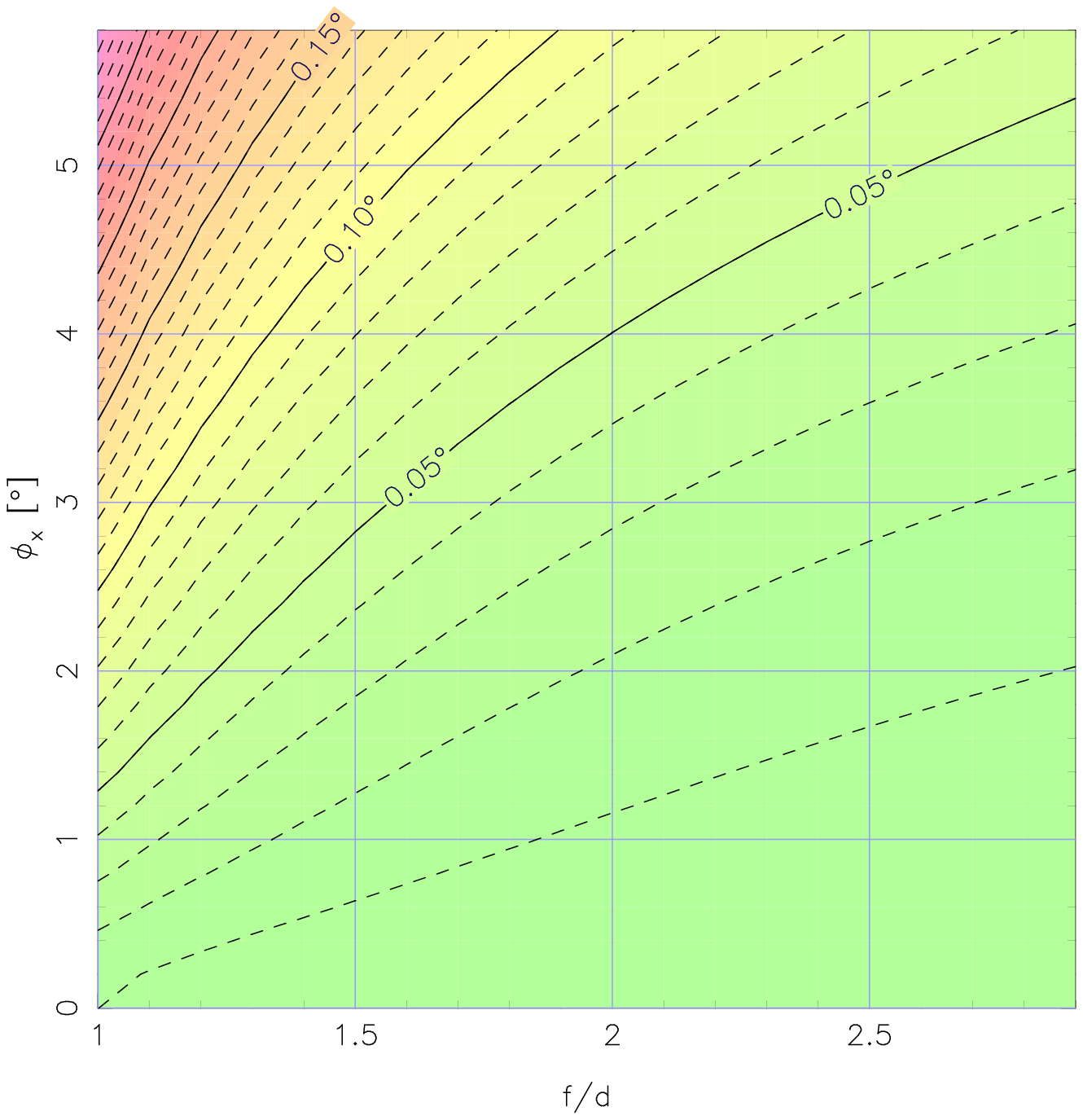}
  }
  \subfigure[
    sagittal rms
  ]{
    \label{f:d-rms-tan}
    \includegraphics[width=0.47\linewidth]{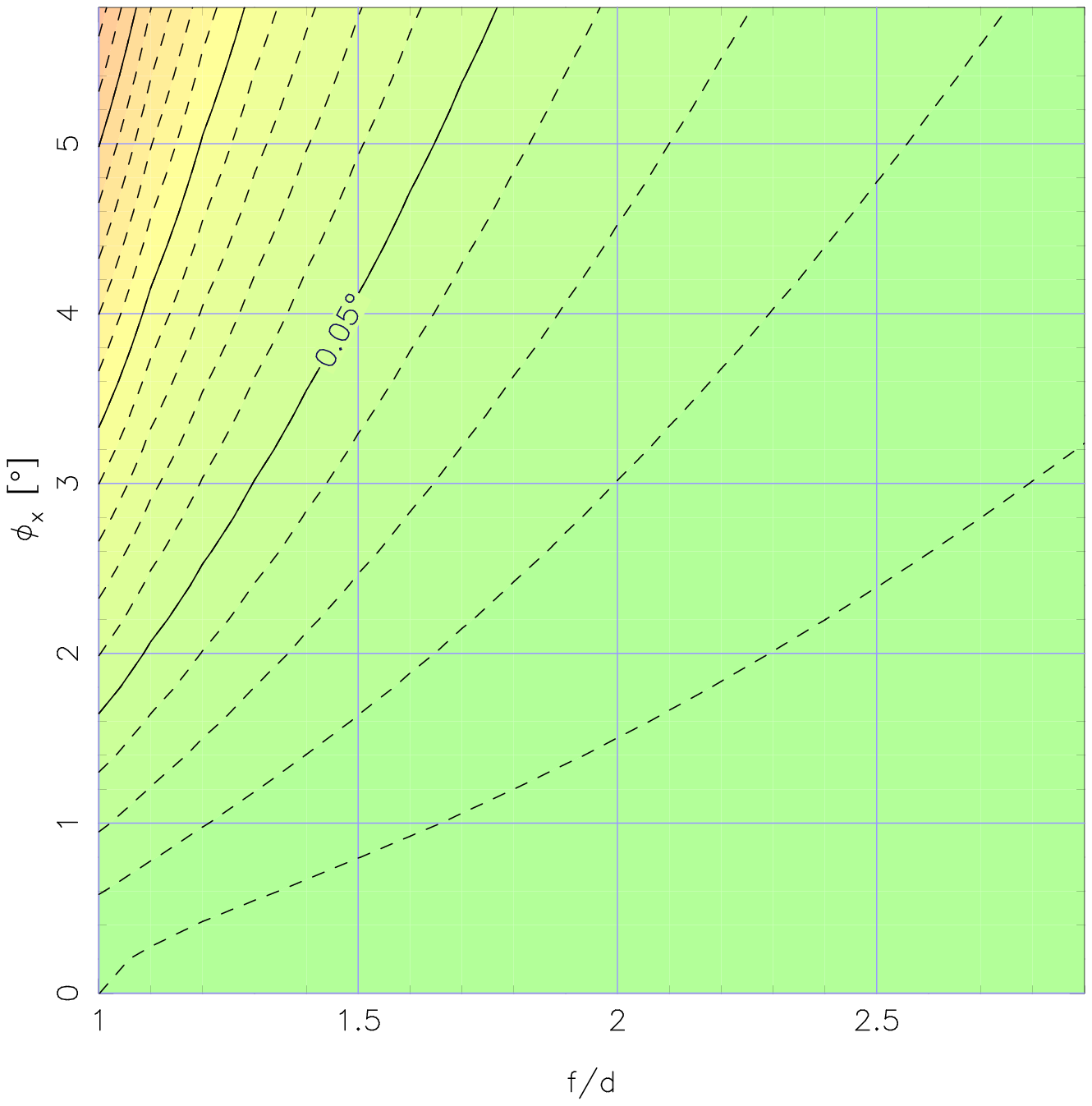}
  }
  \caption{Rms point spread for a \DC{} reflector.
	   The tessellation ratio $\alpha$ is 0.03.
            Illustration analogous to \figref{f:ss-rms}.}
   \label{f:d-rms}
\end{figure}

Another tessellated reflector design being applied in some
of today's IACTs originally goes back to
a solar concentrator and is termed \DC{}
design \cite{Davies1957}.
In this design, the spherical mirror elements are arranged on a 
spheroid with the radius being just the focal length
 of the telescope.
The radius of curvature of the individual mirrors is
constantly $2 f$.
The normals of the mirrors do not coincide with the normals of
the gross spheroid (radius $f$), instead they all point to 
$(x,y,z)=(0,0,2 f)$.
For on-axis incidence, the chief rays of
the single mirrors are imaged perfectly into the focal
point like in the case of a tessellated parabolic reflector \cite{Lewis1990}.
The results for the \DC{} design (\figref{f:d-rms})
exhibit a striking similarity to the data for
a parabolic reflector, although its gross shape is spherical.
For larger incidence angles, this design outperforms the parabolic
configuration.

\subsubsection{Tessellated parabolic design with adjusted radii of curvature}

The last presented design uses a parabolic gross shape which defines
the positions and the orientation of the mirrors. Though, in contrast
to the second discussed scenario, the radii of curvature of the mirrors
are adapted to their varying distance to the focal point $(x,y,z)=(0,0,f)$
in order to avoid defocus aberration of the individual mirror images.

\begin{figure}[p]
  \centering
  \subfigure[
    tangential rms
  ]{
    \label{f:p1-rms-sag}
    \includegraphics[width=0.47\linewidth]{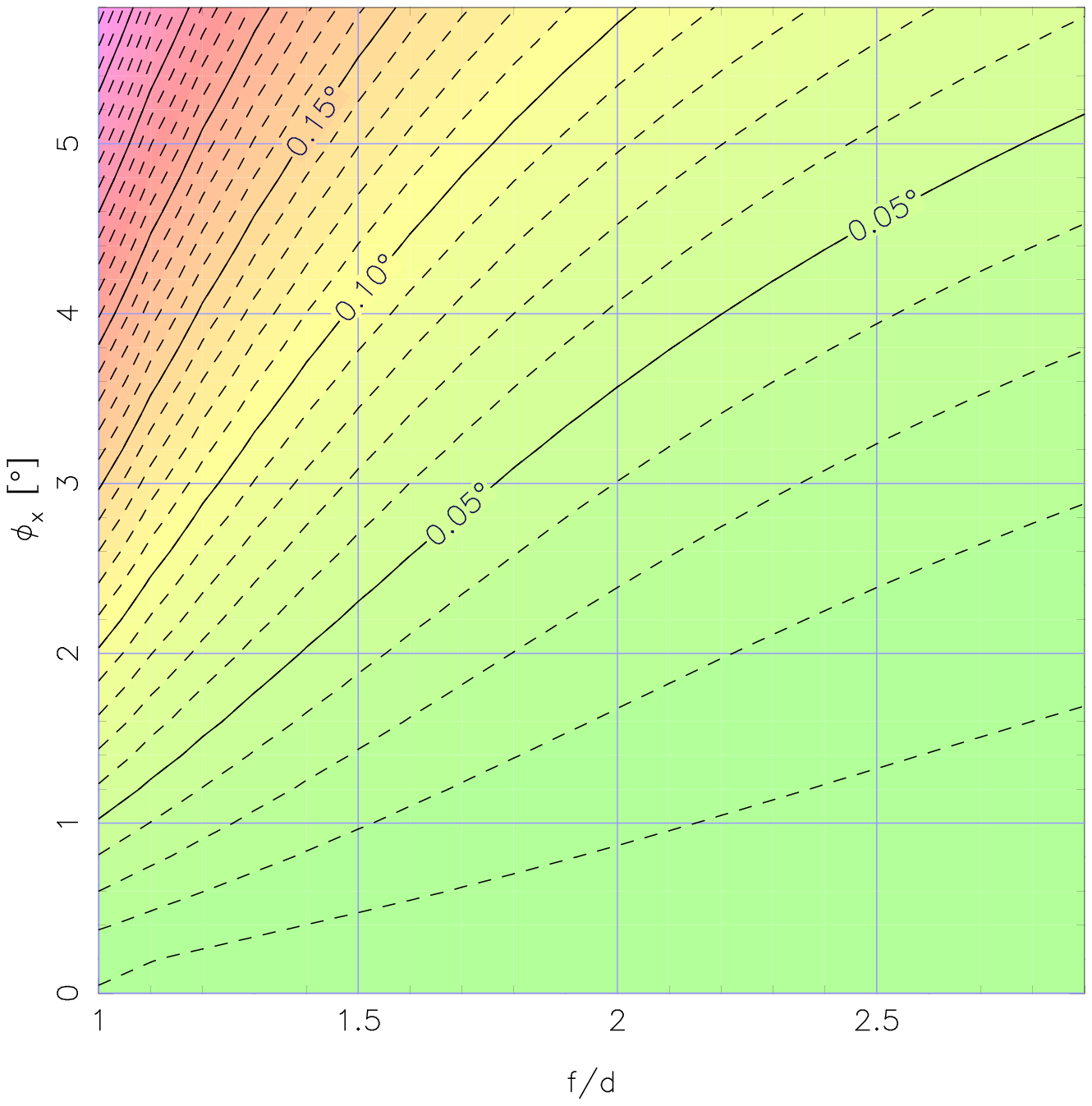}
  }
  \subfigure[
    sagittal rms
  ]{
    \label{f:p1-rms-tan}
    \includegraphics[width=0.47\linewidth]{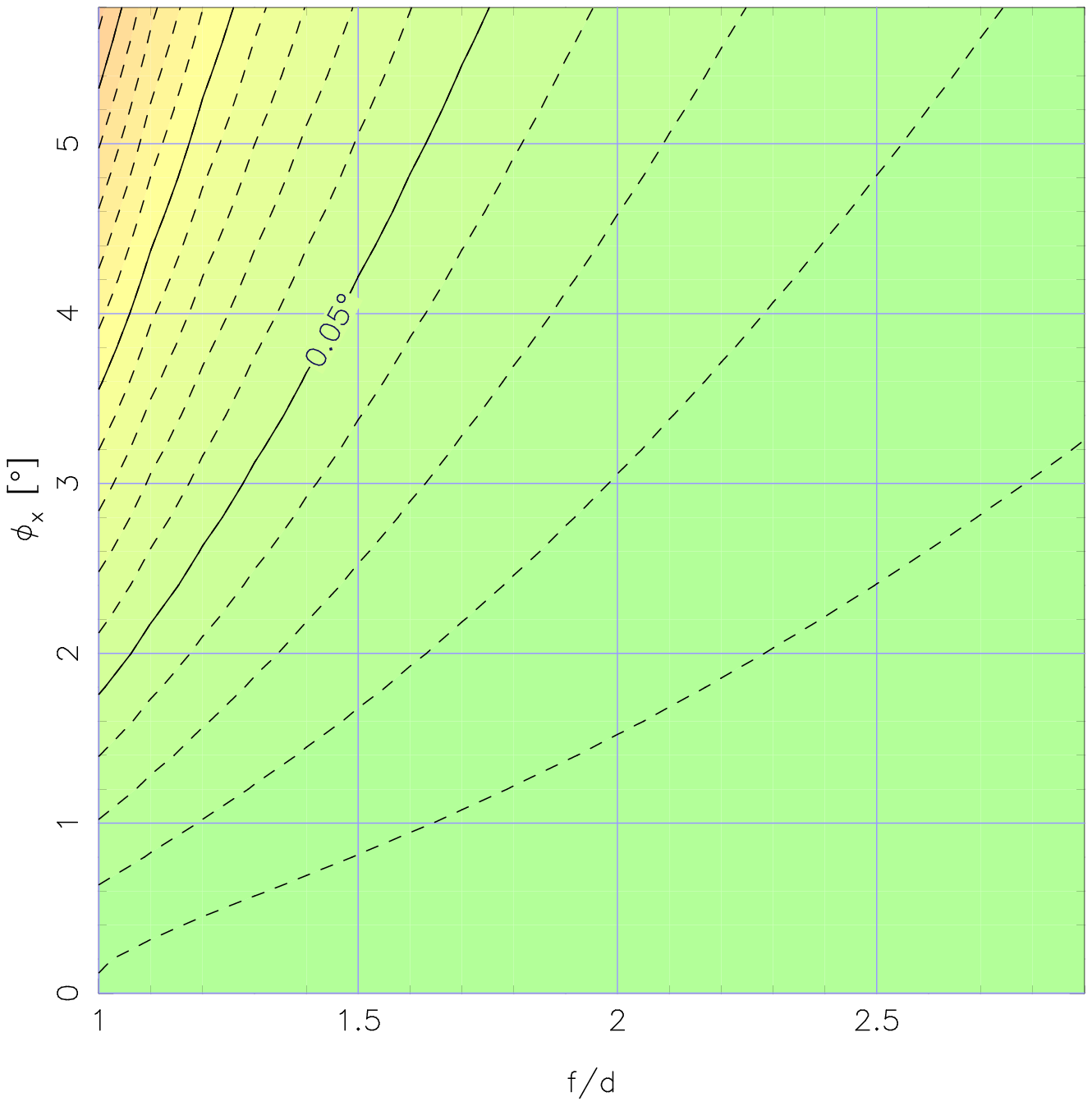}
  }
  \caption{Rms point spread for a tessellated parabolic reflector with adjusted radii.
 	   The tessellation ratio $\alpha$ is 0.03.
 	 Illustration analogous to \figref{f:ss-rms}.}
   \label{f:p1-rms}
\end{figure}

A parabola of revolution has two principal radii in every point
of its surface. As a first approximation, one may take
the average of these two as the radius of curvature of
a mirror segment. Yet, it turns out that especially
for larger distances from the telescope axis, this is
not an ideal solution.
Superior choices of the radii have been found in numerical
optimization runs \cite{Fernandez1998}, \cite{Tonello2002}.
Interpolating and scaling the optimized radii for a parabolic
gross shape from ref.\ \cite{Fernandez1998}, we obtain improved performance
for small incidence angles, as the simulation data show
(\figref{f:p1-rms}).
Yet, for larger angles global comatic aberrations dominate --
just as in the case of the tessellated parabolic reflector
without radius adjustment of the single mirrors.

\section{Discussion}

\subsection{Shower discrimination capability}
The crucial criterion for an optical layout suitable for IACTs
is the ability to discriminate showers induced by hadrons from
those induced by $\gamma$-rays.
In order to enable shower discrimination over a given FOV, a large enough focal ratio has to be chosen
such that both tangential and sagittal rms spot sizes are below
$0{.}05\degree$.

\begin{figure}[htbp]
  \centering
  \includegraphics[width=0.8\linewidth]{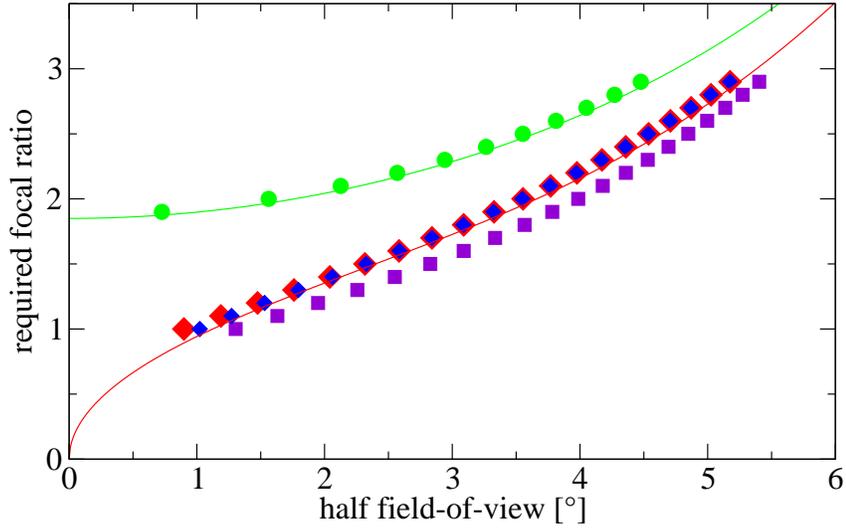}
  \caption{%
  Required focal ratio to distinguish $\gamma$-ray from hadron induced showers
  over a half FOV $\phi$.
  Points: simulation data for
  spherical design (green),
  parabolic design with constant radii (red),
  \DC{} design (violet),
  parabolic design with adjusted radii (blue).
  Tessellation ratio $\alpha$ is 0.03.
  Lines: third-order approximation for
  single-piece paraboloid (red),
  single-piece sphere (green).}
  \label{f:minfod}
\end{figure}

\figref{f:minfod} shows the minimum required focal ratios
for the four presented tessellated reflector designs.
The results have been obtained from the simulation data
for the tangential rms, which is larger than
the sagittal rms in all considered cases.
Although third-order optical theory only treats
single-piece reflectors, the behavior of the systems
can be well predicted.
Solving \eqref{e:torms} for the focal ratio,
good approximations for the results found in the simulations 
can be obtained.
Only the \DC{} data cannot be reproduced
that accurate -- this is because it is the only discussed
design in which the normals (in the centers) of the individual mirrors do
not coincide with the normals of the gross shape;
 a situation that cannot be captured by conventional third-order
analysis.
Yet, qualitatively, the behavior resembles much that of a parabolic 
reflector.

The \DC{} design is superior to all other presented designs:
It allows to make reflectors $\sim 0.2$ faster for all analyzed fields
compared to parabolic gross shapes.
Adjusting the radius of curvature of the individual
mirrors in a parabolic design is only effective when small FOV
($\phi<1.5\degree$) are desired.
Spherical configurations yield the largest spot sizes and, consequently,
only poor shower discrimination capability.

\subsection{Alternative configurations}

The \DC{} design has no single-piece analogue but makes explicit use of
the  new degree of freedom introduced by tessellation.
It is an interesting question whether there are other tessellated 
prime-focus systems with even wider FOV.

\begin{figure}[htb]
  \centering
  \includegraphics[width=0.7\linewidth]{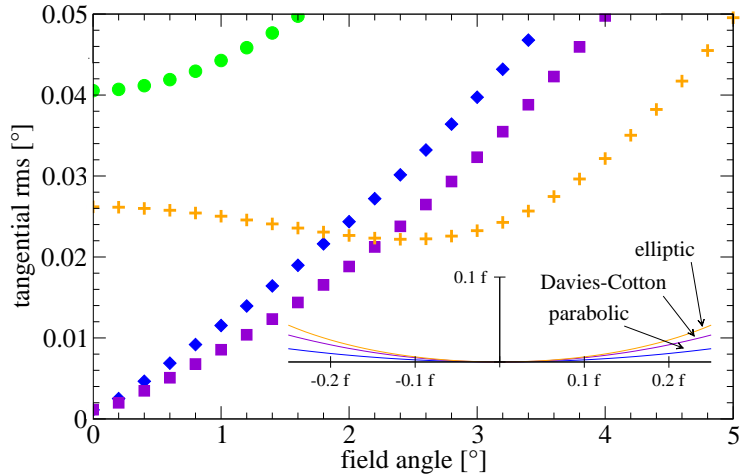}
  \caption{Simulated tangential rms for given field angle
     $\phi$ for
    spherical design (green),
    \DC{} design (violet),
    parabolic design with adjusted radii (blue),
    elliptical design (orange).
    The focal ratio is $2$, tessellation ratio 0.03.
    The inset shows the actual gross shape of the
    different configurations.}
  \label{f:alt}
\end{figure}

A simple approach was taken to answer this question.
For fixed $f$ and $d$, the parameters $r$ and $\delta$ of the gross reflector
shape were varied.
Reasonable imaging of the whole system was
warranted by orientating the individual mirrors (all of which have
a radius of curvature of $2f$)
so that their normals (at their centers) point to $(0,0,2f)$,
like in a \DC{} design \cite{Lewis1990}.
Quite easily, designs allowing even wider
FOV than conventional \DC{} could be found, with
some dependence on the chosen focal ratio.
For $f/d=2$, \figref{f:alt} shows the example of
an elliptical gross shape ($r=0.85 f$, $\delta=5$) which has
a full FOV of $10\degree$.

Besides spatial resolution, high temporal resolution is important
for an effective background suppression especially when measuring in the
sub-$100\unit{GeV}$ energy regime \cite{Akhperjanian2004}.
This means that the arrival time of Cherenkov photons at the camera
should not depend on the point where they hit the reflector.
Parabolic reflectors are (apart from small effects introduced by tessellation)
{\em isochronous}, whereas in a \DC{} design there is a spread in photon
arrival time \cite{Mirzoyan1996}.
We have simulated the photon arrival time distribution for parabolic, \DC{}
and the described elliptic design in the limit $\alpha\rightarrow 0$ and summarize
the main results in \tabref{t:alt}.
The improved off-axis imaging of the elliptic design comes at the expense
of timing accuracy.
The differences in the photon arrival time distributions for different incidence
angles $<5\degree$ were negligible in all considered designs.

\subsection{Tessellation ratio}

\begin{figure}[htb]
  \centering
  \includegraphics[width=0.7\linewidth]{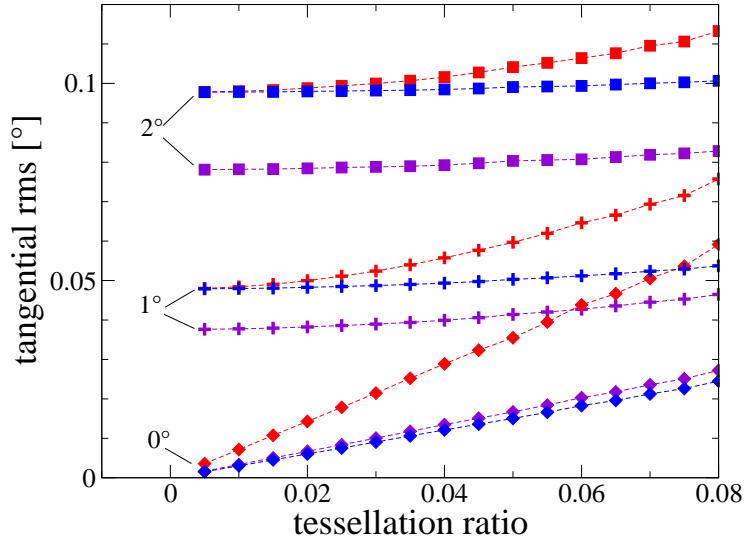}
  \caption{%
  Dependence of tangential rms on tessellation ratio $\alpha$ for
  incidence angles $0\degree$ (diamonds), $1\degree$ (crosses) and $2\degree$ (squares).
  Parabolic design with constant radii (red),
  \DC{} design (violet),
  parabolic design (blue) with adjusted radii.
  Points are connected to guide the eye.
  }
  \label{f:tr}
\end{figure}

The data presented so far were for the fixed tessellation ratio
of $\alpha=0.03$, which corresponds approximately to the value
of the MAGIC telescope, which has a diameter\footnote{Actually,
the MAGIC reflector is an octagon.} of $17\unit{m}$ 
and the individual mirrors are $0.5\unit m\times 0.5 \unit{m}$.
Other ratios are imaginable and may also be desirable for
cost reasons. 
As stated above, all simulations were performed for a range
of tessellation ratios.
Depending on the system configuration, the focal
ratio and incidence angle, it may have
influence on image quality within the investigated range.
The data for a focal ratio of 1,
and incidence angles $0\degree$, $1\degree$ and $2\degree$
are shown in \figref{f:tr}.
The choice of these parameters has practical reasons:
Focal ratios as fast as 1 or slightly more are common in today's
IACTs, and the incidence angles are limited to below $3\degree$.

The parabolic design shows the strongest dependence
on the tessellation ratio.
This is due to the defocus of the individual mirrors that worsens
for larger segments.
If defocus is eliminated by adjusting the radius of curvature
in the parabolic design,
the tessellation ratio is much less critical.
Only for small incidence angles, when global aberrations vanish,
it affects quality distinctly.
As expected, the values of the parabolic configurations converge
in the limit $\alpha\rightarrow 0$.
For the spherical design, the tessellation ratio does not
influence image quality, since the resulting shape is always
the same -- that of a solid spheroid -- independent of
segmentation.
The aberrations are too large to fit in the range depicted
in \figref{f:tr}, though.

Obviously, tessellation ratio
does not deteriorate imaging quality critically as
long as it is below $0.08$ (and the individual radii of
curvature are adjusted in the parabolic case).
Thus, for example, the mirrors of the MAGIC telescope
could have twice the size (then $\alpha\approx 0.06$)
without worsening its performance.

\section{Conclusions}

The main purpose of this paper was to study the potential
of different prime focus designs for wide-angle IACTs, using
third-order optical aberration theory and ray-tracing simulations.
The investigations comprise practically the entire reasonable parameter
range for both single-piece and tessellated parabolic and spherical,
as well as the \DC{} design.
Along with that, some new tessellated designs have been examined.
The \DC{} design exhibits best off-axis performance of the conventional
designs.
Yet, tessellated designs with elliptic gross shapes can yield even wider
FOV but at the expense of timing accuracy.
We show that $f/2$ designs can provide $10\degree$ full FOV.
For faster $f/1$ optics the full FOV available at a $0{.}1\degree$ resolution
is below $3\degree$.
The simulation results also show that for wide-angle IACTs,
segmentation of the gross shape into spherical mirrors deteriorates
imaging only negligibly.

We are planning to study more complex systems that may provide even wider
fields-of-view.

\begin{table}[p]
\centering
\begin{tabular}
  {|c|r|r|r|r|r|}
  \hline\hline
   & $r$ & $\delta$ & $\phi_\stt{max}$ &
   					\begin{minipage}[t]{2 cm}
					  $t_\stt{FWHM}/f$ \\
					  $\phix=0\degree$
					 \end{minipage} &
								\begin{minipage}[t]{2 cm}
								  $t_\stt{FWHM}/f$ \\
								  $\phix=\phi_\stt{max}$
								\end{minipage}\\
  \hline\hline
  parabolic & $2f$      & $-1.0$ & $\sim 3.6\degree$ & $0{.}00\unit{ns}/\unit{m}$&$0{.}00\unit{ns}/\unit{m}$\\
  \DC{}       & $f$       &  $0.0$ & $\sim 4.0\degree$ & $0{.}11\unit{ns}/\unit{m}$& $0{.}11\unit{ns}/\unit{m}$\\
  elliptic  & $0.85f$   & $+5.0$ & $\sim 5.0\degree$ & $0{.}18\unit{ns}/\unit{m}$&$0{.}18\unit{ns}/\unit{m}$\\
  \hline\hline
\end{tabular}
\caption{Comparison of point spread and timing properties
  of some tessellated designs with a gross shape described by
  the radius of curvature $r$ and conic constant $\delta$.
  The focal ratio is $f/d=2$, the tessellation ratio $\alpha=0.03$
  for all systems.
$\phi_\stt{max}$ is the maximum available half field angle
and $t_\stt{FWHM}$ the full width at half maximum of the
photon-arrival time distribution, neglecting tessellation 
($\alpha\rightarrow 0$).
Since $t_\stt{FWHM}$ scales linearly with the dimensions of the system for fixed
$f/d$, it is given normalized to the focal distance.}
\label{t:alt}
\end{table}

\begin{ack}
We gratefully acknowledge support from M.\ Teshima.
\end{ack}

\end{document}